\documentclass[prc,twocolumn,showpacs]{revtex4-1}

\usepackage{enumerate}

\usepackage{booktabs}
\usepackage{color}
\usepackage{graphicx}

\usepackage{amsmath}
\usepackage{amssymb}
\usepackage{times}

\def\bvec#1{\mbox{\boldmath $#1$}}

\newcommand{\what}[1]{\widehat{#1}}

\newcommand{\del}[2]{\frac{\partial #1}{\partial #2}}

\newcommand{\bra}{\langle}
\newcommand{\ket}{\rangle}

\newcommand{\idot}{\cdot}

\newcommand{\beq}{\begin{equation}}
\newcommand{\eeq}{\end{equation}}
\newcommand{\bea}{\begin{eqnarray}}
\newcommand{\eea}{\end{eqnarray}}

\def\fun#1#2{\lower3.6pt\vbox{\baselineskip0pt\lineskip.9pt
 \ialign{$\mathsurround=0pt#1\hfil##\hfil$\crcr#2\crcr\sim\crcr}}}

\begin{document}

\title{
  Pairing effects on vorticity of incident neutron current at quasiparticle resonance energies in $n$-$A$ elastic scattering
}

\author{K. Mizuyama$^{1,2}$, H. Dai Nghia$^{3,4}$, T. Dieu Thuy$^{3}$, N. Hoang Tung$^{5,6}$, T. V. Nhan Hao$^{3,4}$}
\email{corresponding author: tvnhao@hueuni.edu.vn}

\affiliation{
  \textsuperscript{1}
  Institute of Research and Development, Duy Tan University,
  Da Nang 550000, Vietnam
  \\
  \textsuperscript{2}
  Faculty of Natural Sciences,  Duy Tan University, Da Nang 550000, Vietnam
  \\
  \textsuperscript{3}
  Faculty of Physics, University of Education, Hue University,
  34 Le Loi Street, Hue City, Vietnam
  \\
  \textsuperscript{4} Center for Theoretical and Computational Physics,
  University of Education, Hue University, 34 Le Loi Street, Hue City, Vietnam
  \\
  \textsuperscript{5} Department of Nuclear Physics and Nuclear Engineering, Faculty of Physics and Engineering Physics,
University of Science, Ho Chi Minh City 700000, Vietnam
  \\ 
   \textsuperscript{6} Vietnam National University, Ho Chi Minh City 700000, Vietnam
}

\date{\today}

\begin{abstract}
  In this study, we analyzed how the incident neutron current is affected by
  the pairing effect in the neutron-nucleus scattering described within the
  framework of Hartree-Fock-Bogoliubov theory by performing numerical calculations
  in terms of current, vorticity, and circulation of the incident neutron current.
  We found that the pairing effect on the incident neutron flux is completely
  different between particle-type and hole-type quasiparticle resonances.
  In the case of h-type quasiparticle resonance, the pairing acts to prevent the neutron flux
  from entering the nucleus, reducing circulation.
  In the case of p-type quasiparticle resonance, pairing acts to reduce circulation at energies
  lower than the resonance energy, but at energies higher than the resonance energy, the effect
  of pairing on the neutron flux is reversed and, conversely, circulation is increased.
\end{abstract}

\maketitle

\section{Introduction}

Quantum vortices have been introduced as quantized circulation in
superfluids and magnetic flux in superconductors to understand the properties of superfluid
helium, type II superconductors, Bose-Einstein condensation of ultracold atoms,
etc.~\cite{feynmann,Abrikosov,Kennedy,Lin,Stringari,Salomaa},
and have been successfully observed experimentally~\cite{Matthews}.
Superfluid phase transition is thought to be caused by the excitation of quantum vortices.
Also in nuclear physics, the coupling rotation of the deformed nucleus and the intrinsic
vorticity has been discussed as the superfluidity phenomena
~\cite{Mikhailov, Mikhailov2, Brink}.
The effect of superfluid vortices in the interior of neutron stars has also been discussed
~\cite{Elgaroy,Sauls,Baym,Langlois}.

Even if we consider it apart from the nature of superfluidity, a vortex is a phenomenon that
is easy to imagine intuitively and is very characteristic in physics in general, as seen in
typhoons and tornadoes.
In nuclear physics, vortex motion has often been proposed as one of the characteristic
collective modes of nuclei (toroidal mode)~\cite{Semenko,Ryezayeva}.
This is because vortex motion has a unique topological structure and dynamic stability
(as seen in Kelvin's circulation theorem, etc.), and can appear anywhere in a system
where there is a current.

Within the framework of Hartree-Fock-Bogoliubov (HFB) theory, the resonances appearing in
neutron-nucleus ($n$-$A$) scattering
include shape resonances formed by centrifugal barriers and quasiparticle resonances formed
by pair-correlation effects. There are two types of quasiparticle resonances: particle and
hole type quasiparticle resonances~\cite{dobac,kobayashi,jost-hfb,classres}.
In \cite{jost-hfb,classres}, we discussed the formation conditions of the resonances
appearing in $n$-$A$ scattering and the effects of pair correlation in terms of the
S- and K- matrix poles.
It is confirmed that the wave functions of all types of resonances have metastable structures,
except when they are affected by Fano effects~\cite{jost-fano} or correlations from independent
K-matrix poles.
Since the width of the resonant state can be understood as the inverse of the lifetime, it is
consistent that the wave function of the resonant state has a metastable structure and is
observed as a peak with a finite width in the cross section.

In $n$-$A$ scattering, resonances are observed as sharp peaks with small widths at all
cross sections given as a function of incident neutron energy. It is known that the cross section
satisfies the optical theorem when there is no absorption effect on the potential in the
fundamental equations such as the Schr\"{o}dinger equation and the HFB equation,
which means that
the neutron flux current satisfies the continuity equation, i.e., the current is conserved.
According to Helmholtz's theorem, the current can be divided into the current with vortices and
the current without vortices. Regardless of the conservation laws of the current, vortices can
generally always be present in the current, unless the wavefunction (as the solution used to
define the current) is constrained to be vortex-free, or the system requires a vortex-free
condition~\cite{sasaki}.

Given the stability of vortices due to their special topological properties and the metastable
fundamental nature of resonance, the existence and the contribution of vortices to the resonant
state can be expected.
The vortices (vortex-like current) that can appear in the neutron flux of $n$-$A$
scattering may be different from what has been defined as a quantum vortex (whose circulation
can be quantized), but may characterize the resonances in this system.

In this paper, we focus on the vortex (vortex-like current) of neutron flux in $n$-$A$
scattering within the framework of HFB theory, and analyze its relation to resonance. Especially
for the quasiparticle resonances, the relation with pairing correlations will be discussed.

\section{Theoretical and calculation background}

In this paper, the HFB equation is solved by assuming the spherical symmetry.
We adopt the Woods-Saxon potential for the mean-field potential $U$
and pair potential $\Delta$ with same parameters as in Ref.\cite{jost-hfb}.
The chemical potential $\lambda=-1$ MeV  is adopted in order to
set the neutron-rich target of the system.

The neutron current $\bvec{j}$ of the $n$-$A$ scattering system
is defined as
\begin{eqnarray}
  \bvec{j}(\bvec{r})
  =
  \mbox{ Im }
  \psi^{(+)*}_1(\bvec{r})
  \nabla
  \psi^{(+)}_1(\bvec{r})
  \label{def-current}
\end{eqnarray}
by using the upper component of the
HFB scattering wave function $\psi^{(+)}_1(\bvec{r})$.
The vorticity $\bvec{\omega}$ for the neutron current is defined by
\begin{eqnarray}
  \bvec{\omega}(\bvec{r})
  &=&
  \nabla\times\bvec{j}(\bvec{r}).
  \label{def-vort}
  \\
  &=&
  \frac{1}{i}
  \nabla
  \psi^{(+)*}_1(\bvec{r})
  \times
  \nabla
  \psi^{(+)}_1(\bvec{r})
  \label{def-vort2}
\end{eqnarray}
The vorticity is a pseudovector which describes the strength and direction
of the local spinning motion.
The sign of the vorticity is defined as positive for counterclockwise
spinning motion.
The circulation $\Gamma$ which is defined by
\begin{eqnarray}
  \Gamma
  &=&
  \int_S d\bvec{S}
  \idot
  \bvec{\omega}(\bvec{r})
  \label{def-circ}
\end{eqnarray}
is the total amount of vorticity in the area ($S$) enclosed by the closed path.
By the Stokes' theorem, the circulation can be rewritten as
\begin{eqnarray}
  \Gamma
  &=&
  \oint_C d\bvec{l}
  \idot
  \bvec{j}(\bvec{r})
  \label{def-circ2}
\end{eqnarray}
where $C$ represents the closed path to give the area $S$.

In this paper, the direction of the incident neutron is taken as the $z$-axis in the system of n-A scattering (i.e. $\bvec{k}$
of $\exp(i\bvec{k}\idot\bvec{r})$ is chosen as $\bvec{k}=k\bvec{e}_z$).
Since the scattering behavior of neutrons by spherical nuclei should be axially symmetric, we show the numerical results
of the current ($\bvec{j}$) and vorticity ($\bvec{\omega}$) in the $z$-$x$ plane, which is defined by setting $\varphi=0$ in spherical coordinates.
To correctly represent a plane wave $\exp(i\bvec{k}\idot\bvec{r})$ up to 20 MeV in
a region of 20 fm dynamic diameter, $l_{max}=20$ is adopted for the representation of the
partial wave components.

In the $z$-$x$ plane, the vorticity is represented as
\begin{eqnarray}
  \bvec{\omega}(\bvec{r})
  &=&
  2
  \bvec{e}_y
  \mbox{ Im }
  \left[
    (\bvec{e}_z\idot\nabla\psi_1^{(+)*}(\bvec{r}))
    (\bvec{e}_x\idot\nabla\psi_1^{(+)}(\bvec{r}))
    \right]_{\varphi=0}
\end{eqnarray}
where $\bvec{e}_x$, $\bvec{e}_y$ and $\bvec{e}_z$ are the unit vectors for $x$-, $y$- and $z$-axis,
respectively. Using the partial components of the scattering wave function,
$\bvec{e}_z\idot\nabla\psi_1^{(+)}(\bvec{r})$ and $\bvec{e}_x\idot\nabla\psi_1^{(+)}(\bvec{r})$ are
represented as
\begin{eqnarray}
  &&
  \bvec{e}_{z}\idot
  \nabla\psi_1^{(+)}(\bvec{r})
  \nonumber\\
  &&=
  \frac{1}{r}
  \sum_{lj}
  (i)^l
  \frac{2j+1}{2}
  \frac{\sqrt{4\pi}}{2l+1}
  \nonumber\\
  &&\times
  \left[
    \frac{l+1}{\sqrt{2l+3}}
    Y_{l+1,0}(\what{\bvec{r}})
    \left(
    \del{}{r}
    -
    \frac{l+1}{r}
    \right)
    \right.
    \nonumber\\
    &&\hspace{1cm}
    +
    \left.
    \frac{l}{\sqrt{2l-1}}
    Y_{l-1,0}(\what{\bvec{r}})
    \left(
    \del{}{r}
    +
    \frac{l}{r}
    \right)
    \right]
  \psi_{1,lj}^{(+)}(r)
  \label{nablaz}
  \\
  &&
  \bvec{e}_{x}\idot
  \nabla\psi_1^{(+)}(\bvec{r})
  \nonumber\\
  &&=
  \frac{1}{r}
  \sum_{lj}
  (i)^l
  \frac{2j+1}{2}
  \frac{\sqrt{4\pi}}{2l+1}
  \nonumber\\
  &&\times
  \left[
    -
    \sqrt{\frac{(l+1)(l+2)}{2l+3}}
    Y_{l+1,1}(\what{\bvec{r}})
    \left(
    \del{}{r}
    -
    \frac{l+1}{r}
    \right)
    \right.
    \nonumber\\
    &&\hspace{1cm}
    +
    \left.
    \sqrt{\frac{l(l-1)}{2l-1}}
    Y_{l-1,1}(\what{\bvec{r}})
    \left(
    \del{}{r}
    +
    \frac{l}{r}
    \right)
    \right]
  \psi_{1,lj}^{(+)}(r).
  \label{nablax}
\end{eqnarray}
Note that the spin non-flip is supposed in these expressions.


We can obtain
the Lippmann-Schwinger (LS) integral equation for the HFB scattering wave function
\begin{eqnarray}
  &&
  \begin{pmatrix}
    \psi_1^{(+)}(\bvec{r}) \\
    \psi_2^{(+)}(\bvec{r})
  \end{pmatrix}
  =
  \begin{pmatrix}
    \psi_0^{(+)}(\bvec{r}) \\
    0
  \end{pmatrix}
  \nonumber\\
  &&
  +
  \int d\bvec{r}_1
  \begin{pmatrix}
    G_{0}^{(+)}(\bvec{r},\bvec{r}_1;k_1(E)) &
    0 \\
    0 &
    -G_{0}^{(+)}(\bvec{r},\bvec{r}_1;k_2(E))
  \end{pmatrix}
  \nonumber\\
  &&\times
  \begin{pmatrix}
    0 & \Delta(\bvec{r}_1) \\
    \Delta(\bvec{r}_1) & 0
  \end{pmatrix}
  \begin{pmatrix}
    \psi_1^{(+)}(\bvec{r}_1) \\
    \psi_2^{(+)}(\bvec{r}_1)
  \end{pmatrix}
  \label{LSHFB}
\end{eqnarray}
by applying the two-potential formula, where $\psi_0^{(+)}$ is the Hartree-Fock (HF) scattering wave function
and the HF Green's function which satisfies
\begin{eqnarray}
  \left(\frac{\hbar^2k^2}{2m}+\frac{\hbar^2}{2m}\nabla^2-U\right)
  G_{0}^{(+)}(\bvec{r},\bvec{r}';k)=\delta(\bvec{r}-\bvec{r}'),
\end{eqnarray}
$k_1(E)$ and $k_2(E)$ are defined by $k_1(E)=\sqrt{\frac{2m}{\hbar^2}(\lambda+E)}$ and
$k_2(E)=\sqrt{\frac{2m}{\hbar^2}(\lambda-E)}$ as is introduced in Ref.\cite{jost-hfb}.

The LS equation Eq.(\ref{LSHFB}) can be expressed in the form of independent integral equations for the upper
and lower components, respectively, using the HFB Green's function $\mathcal{G}$ and the Dyson equation that it satisfies.

The upper component can be expressed in two ways as
\begin{eqnarray}
  &&
  \psi_1^{(+)}(\bvec{r})
  \nonumber\\
  &&
  =
  \left\{
  \begin{array}{l}
    \psi_0^{(+)}(\bvec{r})
    \\
    +
    \displaystyle{
    \int\int d\bvec{r}_1d\bvec{r}_2
    \mathcal{G}_{11}^{(+)}(\bvec{r},\bvec{r}_1)
    \Sigma_0(\bvec{r}_1,\bvec{r}_2)
    \psi_0^{(+)}(\bvec{r}_2)}
    \\
    \\
    \psi_0^{(+)}(\bvec{r})
    \\
    +
    \displaystyle{
    \int\int d\bvec{r}_1d\bvec{r}_2
    G_{0}^{(+)}(\bvec{r},\bvec{r}_1;k_1)
    \Sigma(\bvec{r}_1,\bvec{r}_2)
    \psi_0^{(+)}(\bvec{r}_2)},
  \end{array}
  \right.
  \label{psi1}
\end{eqnarray}
and the lower component is expressed in a form that is computed using
the upper component as
\begin{eqnarray}
  &&
  \psi_2^{(+)}(\bvec{r})
  =
  -
  \int d\bvec{r}_1
  G_{0}^{(+)}(\bvec{r},\bvec{r}_1;k_2)
  \Delta(\bvec{r}_1)
  \psi_1^{(+)}(\bvec{r}_1)
  \label{psi2},
\end{eqnarray}
with $\Sigma_0$ and $\Sigma$ which are defined by
\begin{eqnarray}
  \Sigma_0(\bvec{r},\bvec{r}')
  &=&
  -\Delta(\bvec{r})
  G_{HF}^{(+)}(\bvec{r},\bvec{r}';k_2)
  \Delta(\bvec{r}')
  \label{Sig0}
  \\
  \Sigma(\bvec{r},\bvec{r}')
  &=&
  \Delta(\bvec{r})
  \mathcal{G}_{22}^{(+)}(\bvec{r},\bvec{r}')
  \Delta(\bvec{r}'),
  \label{Sig}
\end{eqnarray}
where $\mathcal{G}_{11}^{(+)}$ and $\mathcal{G}_{22}^{(+)}$ are the diagonal components of the
HFB Green's function which is expressed in $4\times 4$ matrix form.
It should be noticed that Eq.(\ref{Sig0}) is the first term to appear in the Dyson series expansion
of Eq.(\ref{Sig}) like $\Sigma=\Sigma_0+\cdots$.

Of course, it is obvious that $\Sigma_0$ and $\Sigma$ would behave as the non-local potential
which originates from the pair correlation, and it is important to analyze the role of its non-locality,
but we will not discuss it yet in this paper.

In the HFB theory, it is known that quasiparticle states exist symmetrically in the positive and
negative regions of quasiparticle energy.
This is also true for the resonances (or more precisely, the poles of the S-matrix corresponding
to the resonances)~\cite{jost-hfb}.
There are two types of quasiparticle resonances, particle-type (p-type) and hole-type (h-type), both of which exist
in both the positive and negative quasiparticle energy regions.

By performing the similar approximate calculations to derive the spectral representation of the HFB
Green function as shown in Ref.\cite{jost-fano},
it is possible to confirm that in the positive energy region ($E>0$), a particle-type
resonance behaves as a pole of $\mathcal{G}_{11}$ and a hole-type resonance behaves as a pole of
$\mathcal{G}_{22}$, i.e.
\begin{eqnarray}
  \bra p|\mathcal{G}_{11}|p\ket
  &\sim&
  \frac{1}{E-E_{qp}^{(p)}}
  \label{pres}
  \\
  \bra h|\mathcal{G}_{22}|h\ket
  &\sim&
  \frac{1}{E-E_{qp}^{(h)}}
  \label{hres}
\end{eqnarray}
where $E_{qp}^{(p)}$ and $E_{qp}^{(h)}$ are the S-matrix poles for the particle-type and hole-type quasiparticle
resonance energy which are given as the complex number.
The detail expressions of $E_{qp}^{(p)}$ and $E_{qp}^{(h)}$ are given in Refs.\cite{jost-hfb,classres}.
This is the origin of the metastable structure of the scattering wave function at the resonance energy.

By inserting Eq.(\ref{psi1}) into Eqs.(\ref{def-current}),
we can divide the current $\bvec{j}$ into three terms as
\begin{eqnarray}
  \bvec{j}
  &=&
  \bvec{j}_0
  +
  \bvec{j}_{int}
  +
  \delta\bvec{j}
  \label{jdev}
\end{eqnarray}
with
\begin{eqnarray}
  \bvec{j}_0
  &=&
  \mbox{Im }
  \psi_0^{(+)*}
  \nabla
  \psi_0^{(+)}
  \label{j0}
  \\
  \bvec{j}_{int}
  &=&
  \mbox{Im }
  \left[
  \delta\psi^{(+)*}
  \nabla
  \psi_0^{(+)}
  +
  \psi_0^{(+)*}
  \nabla
  \delta\psi^{(+)}
  \right]
  \label{jint}
  \\
  \delta\bvec{j}
  &=&
  \mbox{Im }
  \delta\psi^{(+)*}
  \nabla
  \delta\psi^{(+)}
  \label{delj}
\end{eqnarray}
where $\delta\psi^{(+)}$ is the second term of Eq.(\ref{psi1})
which includes the pair correlation, $\bvec{j}_0$ is the HF current,
$\bvec{j}_{int}$ is the current which represents the interference
between the HF scattering wavefunction $\psi^{(+)}_{0}$ and $\delta\psi^{(+)}$,
and $\delta\bvec{j}$ is the current formed by the scattered wave by the pair
potential in the HF mean field ($\delta\psi^{(+)}$).

By using Eqs.(\ref{jdev}), the vorticity is also divided into three terms as
\begin{eqnarray}
  \bvec{\omega}
  &=&
  \bvec{\omega}_0
  +
  \bvec{\omega}_{int}
  +
  \delta\bvec{\omega}
\end{eqnarray}
with
\begin{eqnarray}
  \bvec{\omega}_0
  &=&
  \nabla\times\bvec{j}_0
  \nonumber\\
  &=&
  2
  \bvec{e}_y
  \mbox{ Im }
  \left[
    (\bvec{e}_z\idot\nabla\psi^{(+)*}_0)
    (\bvec{e}_x\idot\nabla\psi^{(+)}_0)
    \right]
  \\
  \bvec{\omega}_{int}
  &=&
  \nabla\times\bvec{j}_{int}
  \nonumber\\
  &=&
  2
  \bvec{e}_y
  \mbox{ Im }
  \left[
    (\bvec{e}_z\idot\nabla\delta\psi^{(+)*})
    (\bvec{e}_x\idot\nabla\psi^{(+)}_0)
    \right.
    \nonumber\\
    &&\hspace{10pt}
    -
    \left.
    (\bvec{e}_x\idot\nabla\delta\psi^{(+)*})
    (\bvec{e}_z\idot\nabla\psi^{(+)}_0)
    \right]
  \\
  \delta\bvec{\omega}
  &=&
  \nabla\times\delta\bvec{j}
  \nonumber\\
  &=&
  2
  \bvec{e}_y
  \mbox{ Im }
  \left[
    (\bvec{e}_z\idot\nabla\delta\psi^{(+)*})
    (\bvec{e}_x\idot\nabla\delta\psi^{(+)})
    \right].
  \nonumber\\
\end{eqnarray}
$\delta\bvec{j}$ and $\delta\bvec{\omega}$ are expected to be smaller in absolute value
(than $\bvec{j}_{int}$ and $\bvec{\omega}_{int}$) because they are quantities that
give higher-order pairing contributions. However, the continuity equation for $\delta\bvec{j}$
and $\bvec{j}_{int}$ is given by
\begin{eqnarray}
  \nabla\idot(\bvec{j}_{int}+\delta\bvec{j})=0
  \label{contjint}
\end{eqnarray}
since $\bvec{j}$ and $\bvec{j}_0$ satisfy $\nabla\idot\bvec{j}=0$ and $\nabla\idot\bvec{j}_0=0$,
respectively. Eq.(\ref{contjint}) is shown to say $\nabla\idot\bvec{j}_{int}\neq 0$ as long as
$\delta\bvec{j}\neq 0$, $\nabla\idot\bvec{j}_{int}\neq 0$ implies that $\bvec{j}_{int}$ contains
the ``gushing'' or ``suctioning'' motion of the current.
If we define
\begin{eqnarray}
  \bvec{j}_{pair}
  &=&\bvec{j}_{int}+\delta\bvec{j}
  =\bvec{j}-\bvec{j}_{0}
  \label{jpair}
\end{eqnarray}
as the current represents the pairing contribution,
$\bvec{j}_{pair}$ doesn't include ``gushing'' or ``suctioning'' motion, i.e. $\nabla\idot\bvec{j}_{pair}=0$.
The vorticity which corresponds to $\bvec{j}_{pair}$ ($\bvec{\omega}_{pair}$) is also defined as
\begin{eqnarray}
  \bvec{\omega}_{pair}
  &=&
  \nabla\times\bvec{j}_{pair}
  \\
  &=&
  \bvec{\omega}_{int}+\delta\bvec{\omega}
  =
  \bvec{\omega}-\bvec{\omega}_{0}
\end{eqnarray}

The approximated expression of $\bvec{\omega}_{pair}$ near the quasiparticle resonance energy
can be represented as

\begin{widetext}
\begin{eqnarray}
  \bvec{\omega}_{pair}
  &\sim&
  \left\{
  \begin{array}{l}
    \mbox{[For particle-type quasiparticle resonance]}
    \\
    \displaystyle{
      \frac{(E-\mbox{Re }E_{qp}^{(p)})\mbox{ Re }\bra p|\Sigma_0^{(1)}|\psi^{(+)}_0\ket}
           {(E-\mbox{Re }E_{qp}^{(p)})^2+(\mbox{Im }E_{qp}^{(p)})^2}
      2
      \bvec{e}_y
      \mbox{ Im }
      \left[
        (\bvec{e}_z\idot\nabla\phi_p^*)
        (\bvec{e}_x\idot\nabla\psi^{(+)}_0)
        -
        (\bvec{e}_x\idot\nabla\phi_p^*)
        (\bvec{e}_z\idot\nabla\psi^{(+)}_0)
        \right]
      }
      \\
      +(\mbox{Other partial wave components of } \bvec{\omega}_{int})
      +\delta\bvec{\omega}
    \\
    \\
    \mbox{[For hole-type quasiparticle resonance]}
    \\
    \displaystyle{
      -\frac{(\mbox{Im }E_{qp}^{(h)})^2}{(E-\mbox{Re }E_{qp}^{(h)})^2+(\mbox{Im }E_{qp}^{(h)})^2}
      \bvec{\omega}_{0}
      +(\mbox{Other partial wave components of } \bvec{\omega}_{int})
      +\delta\bvec{\omega}
      }
  \end{array}
  \right.
  \label{approx-omega}
\end{eqnarray}
\end{widetext}
The first term of Eq.(\ref{approx-omega}) can be obtained as a leading order term of $\bvec{\omega}_{int}$
by applying Eqs.(\ref{pres}) and (\ref{hres}) to Eq.(\ref{psi1}).
The contribution of other partial wave components of $\bvec{\omega}_{int}$ and $\delta\bvec{\omega}$
are expected to be very small compared to that of the first term.
From Eq.(\ref{approx-omega}), we can expect that the contribution of pairing to vorticity is quite different
between p-type and h-type quasiparticle resonance. In the case of p-type, the positive and negative contributions
of $\bvec{\omega}_{pair}$ are reversed depending on whether $E$ is greater or less than $E_{qp}$. In the case of
h-type, if we can ignore all but the first term (leading term), the pairing effect leads to zero vorticity,
i.e. the vortices disappear at $E\sim \mbox{ Re }E_{qp}^{(h)}$.

\begin{figure}[htbp]
\includegraphics[width=\linewidth,angle=0]{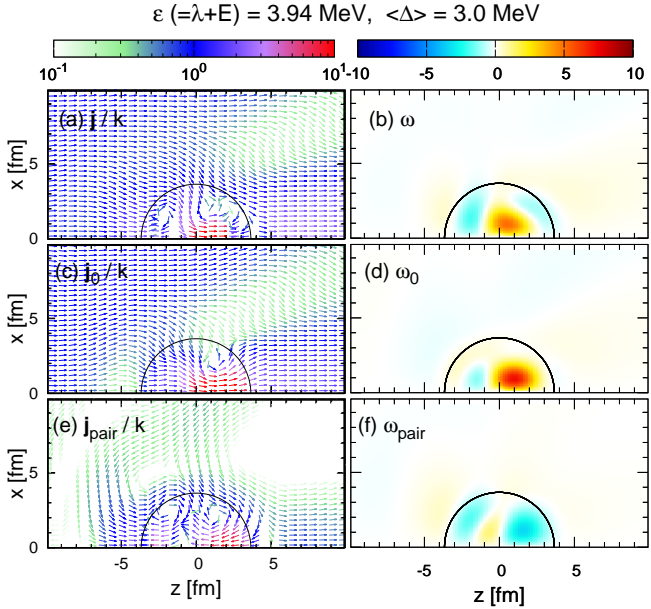}
\caption{(Color online) The neutron currents (a)$\bvec{j}$,
  (c)$\bvec{j}_0$ and (e)$\bvec{j}_{pair}$ at $\epsilon=3.94$ MeV
  (a corresponding energy to the h-type $d_{3/2}$ resonance)
  with $\lambda=-1.0$ MeV and $\bra\Delta\ket=3.0$ MeV.
  The panels (b), (d) and (f) are the corresponding
  vortices to the currents (a), (c) and (e), respectively.
  The arrows length of the currents are represented by the
  same length. The strength (absolute value) of the current
  which is normalized by $k\left(=\sqrt{\frac{2m}{\hbar^2}(\lambda+E)}\right)$
  (the absolute value of the plane wave) is represented by color.
  The unit of the vorticity is fm$^2$. The black solid semicircle shows
  the size of the target nucleus given by $r_0 A^{\frac{1}{3}}=3.66$ fm. }
\label{fig1}
\end{figure}

\section{Numerical results}

In Fig.\ref{fig1}, the neutron currents ($\bvec{j}$, $\bvec{j}_0$, and $\bvec{j}_{pair}$)
and the vorticities ($\bvec{\omega}$, $\bvec{\omega}_0$ and $\bvec{\omega}_{pair}$) at the
energy corresponding to the h-type $d_{3/2}$ resonance are shown in the left and right
panels, respectively.

As given in Eqs.(\ref{def-circ}) and (\ref{def-circ2}),
the circulation ($\Gamma$, $\Gamma_0$ and $\Gamma_{pair}$) are calculated by the
area integration of $\bvec{\omega}$, $\bvec{\omega}_0$ and $\bvec{\omega}_{pair}$
or line integration of $\bvec{j}$, $\bvec{j}_0$, and $\bvec{j}_{pair}$.
Since the current is symmetric about the z-axis, the region of the upper half of
$z$-$x$ plane $0 \leq x \leq 10$ fm, $-10 \leq z \leq 10$ fm (the area showing
current and vorticity in Fig.\ref{fig1})
is adopted as the integration area for the calculation of the circulation.
\begin{figure}[htbp]
\includegraphics[width=\linewidth,angle=0]{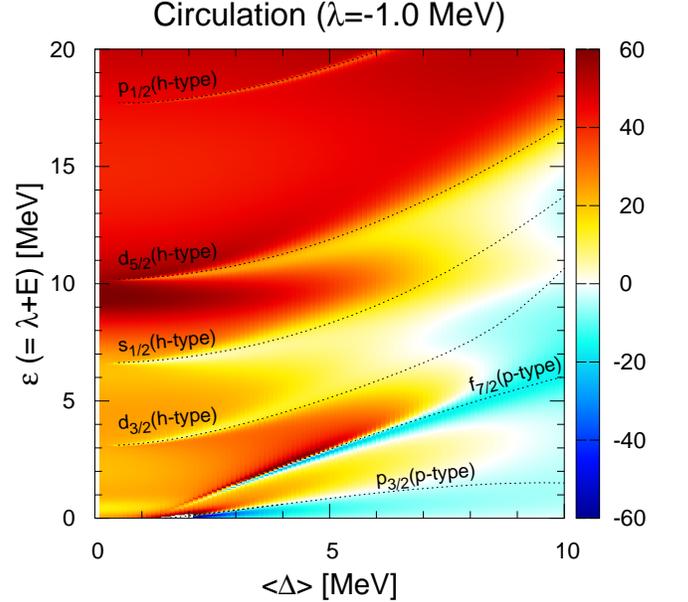}
\caption{(Color online) The contour plot of $\Gamma$
  as a function of $\epsilon$ and $\bra\Delta\ket$.
  The dotted curves represent the real part of the
  S-matrix poles for the quasiparticle resonances
  which were shown in Ref.\cite{jost-hfb}.}
\label{fig2}
\end{figure}

In Fig.\ref{fig2}, the contour plot of the circulation $\Gamma$ is shown as a function
of the energy $\epsilon$ and mean pairing gap $\bra\Delta\ket$. The black dotted curves are
the real part of the S-matrix poles for the quasiparticle resonances shown in Figs.6
and 7 in Ref.\cite{classres}.
The curvilinear pattern in the contour plot of the circulation coincides with the curve
of the real part of the pole of the S-matrix of the quasiparticle resonance, indicating
that the circulation of the incident neutron flux in $n$-$A$ scattering is greatly
affected by the pairing effect at the quasiparticle resonance energy.

\begin{figure}[htbp]
\includegraphics[width=\linewidth,angle=0]{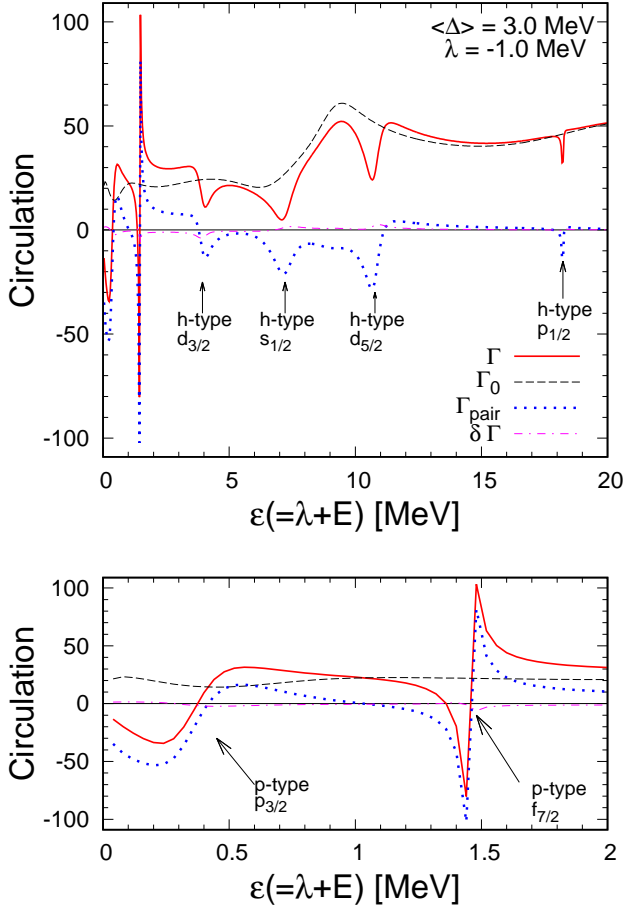}
\caption{(Color online) The circulation obtained by integrating the area of the
  region $0 \leq x \leq 10$ fm, $-10 \leq z \leq 10$ fm (the area as shown in
  Fig.\ref{fig1}), or by line integrating this region in a counterclockwise
  direction. $\Gamma$(red solid), $\Gamma_0$(black dash), $\Gamma_{pair}$(blue dot)
  and $\delta\Gamma$(purple dot-dash) are the circulations which are corresponding to
  $\bvec{\omega}$, $\bvec{\omega}_0$, $\bvec{\omega}_{pair}$ and
  $\delta\bvec{\omega}$, respectively. The lower panel shows an enlargement of the
  region $0\leq\epsilon\leq 2$ MeV to make it easier to see the circulation for the
  p-type resonances.}
\label{fig3}
\end{figure}

In Fig.\ref{fig3}, $\Gamma$, $\Gamma_0$ and $\Gamma_{pair}$ calculated with
$\bra\Delta\ket=3.0$ MeV are plotted as a function of the energy $\epsilon$,
and $\delta\Gamma$ which is obtained from  $\delta\bvec{\omega}$(or $\delta\bvec{j}$)
is also shown together in order to see the contribution.
In the previous section, we expected that $\delta\bvec{\omega}$ would be negligible as a contribution
to $\bvec{\omega}_{pair}$ because it provides a higher-order contribution of the pairing.
The correctness of our expectation is confirmed by the fact that $\delta\Gamma$,
shown as a purple dotted curve in Fig.\ref{fig3}, is actually negligibly small.
The characteristics of the p-type and h-type quasiparticle resonances for $\bvec{\omega}_{pair}$
shown in Eq.(\ref{approx-omega}) are clearly expressed in the energy dependence of the
circulation $\Gamma_{pair}$ in Fig.\ref{fig3}.

As shown in Eq.(\ref{approx-omega}), the $\Gamma_{pair}$ calculated from $\bvec{\omega}_{pair}$
takes the form of a Lorentz distribution and acts to reduce $\Gamma$ near the h-type quasiparticle
resonance.
From the current point of view, it can be seen from Fig.\ref{fig1} that the pairing effect, $\bvec{j}_{pair}$,
occurs in the direction of counteracting the current $\bvec{j}_{0}$ in the nucleus,
and acts to reduce the vorticity $\bvec{\omega}$.
In the case of the h-type quasiparticle resonances, the pairing correlation works to create a flow along
the surface while preventing the incident neutron current from entering the target nucleus.

As indicated in Eq.(\ref{approx-omega}), the effect of pairing on p-type quasiparticle resonances is
quite different. In the case of p-type quasiparticle resonance, the pairing acts to reduce the
circulation $\Gamma$ on the lower side of the resonance energy (the real part of the S-matrix pole),
but on the contrary, it acts to increase the circulation $\Gamma$ in the energy region higher than
the resonance energy.
\begin{figure}[htbp]
\includegraphics[width=\linewidth,angle=0]{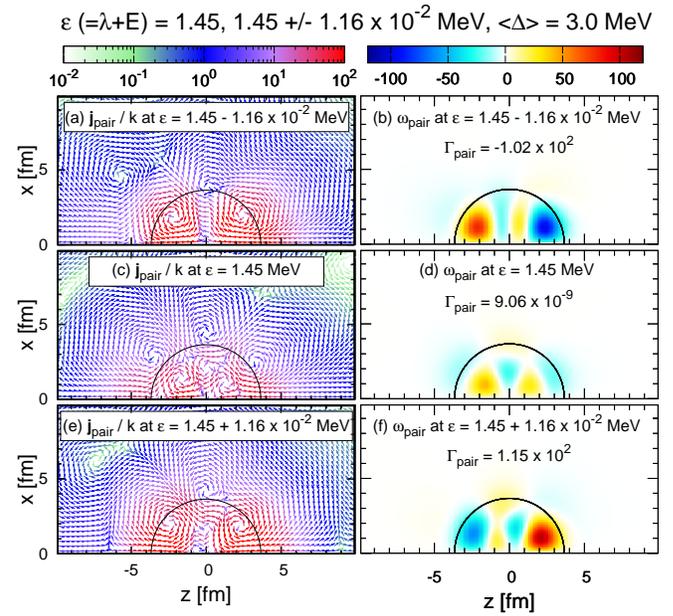}
\caption{(Color online) The current $\bvec{j}_{pair}$ and
  vorticity $\bvec{\omega}_{pair}$ at $\epsilon=1.45-1.16\times 10^{-2}$,
  $1.45$ and $1.45+1.16\times 10^{-2}$ MeV. $\epsilon=1.45$ MeV is the energy
  at which $\Gamma_{pair}$ becomes zero (found near the real part of the
  S-matrix pole of the p-type $f_{7/2}$ resonance), and $\pm 1.16\times 10^{-2}$
  MeV is the imaginary part of the S-matrix pole (half width of the resonance).}
\label{fig4}
\end{figure}
Fig.\ref{fig4} shows the currents $\bvec{j}_{pair}$ and their vorticity $\bvec{\omega}_{pair}$
for the p-type quasiparticle resonance at $f_{7/2}$.
Panels (a), (c), and (e) show the currents at $\epsilon=1.45-1.16\times 10^{-2}$, $1.45$, and
$1.45+1.16\times 10^{-2}$ MeV, respectively, and panels (b), (d), and (f) show the corresponding
vorticity, where $1.45$ MeV is the quasiparticle energy (the real part of the S-matrix pole of
$f_{7/2}$), and $1.16\times 10^{-2}$ MeV is the half width of the resonance (the imaginary part
of the S-matrix pole).

In Fig.\ref{fig4}, we can see the obvious vortex that the current creates. In panel (a),
there is a clockwise vortex in the $z>0$ region and a counterclockwise vortex in the $z<0$ region.
In panel (e), these vortices are reversed. Since the vortices in the $z>0$ region are stronger
in both cases, the vorticity is negative in panel (a) and positive in panel (e).
In panel (c), the weak vortices are symmetrically located inside and outside the nucleus,
so that the overall vorticity is almost zero.
In panel (c), the weak vortices are symmetrically located inside and outside the nucleus, so that
the overall vorticity is almost zero.
It should be noted that, since the current is symmetric about the $z$-axis, the vortex is a
torus-shaped vortex (like a donut about the $z$-axis).

As shown in Ref.\cite{classres}, the $f_{7/2}$ p-type quasiparticle resonance is a resonant state
that appears at energies lower than the centrifugal barrier, and the behavior of the $f_{7/2}$
partial wave component of the scattering HFB wave function is more like a bound state than a
metastable state.
Since $\bvec{j}_{int}$ is dominant in $\bvec{j}_{pair}$ defined in Eq.(\ref{jpair}) and $\bvec{j}_{int}$
is defined in Eq.(\ref{jint}), the vortices arise from the interference between the partial wave
component of the HFB wave function for $f_{7/2}$ and the HF scattering wave function.

By carefully observing the current flow in Fig.\ref{fig4}, we can see that in panel (a), the current is only allowed
to enter the nucleus from the front ($z>0$ region, $\theta\sim 0$), while in panel (e), the current is
allowed to enter from the back region ($z<0$ region).
Qualitatively, this is a common property of p-type quasiparticle resonances, although the angle
at which the current is allowed to flow into the nucleus from back region and the formation of
vortices are different for $p_{3/2}$ and $f_{7/2}$.

In Ref.\cite{classres}, we showed that when the independent K-matrix pole is located near the
quasiparticle resonance, the correlation between the K-matrix pole of the quasiparticle resonance
and the independent K-matrix pole causes the K-matrix pole to disappear when the pairing gap
$\bra\Delta\ket$ becomes larger than the critical gap $\bra\Delta\ket_c$.
We also showed that the critical gap of the p-type $p_{3/2}$ is $\bra\Delta\ket_c=4.08$ MeV.
\begin{figure}[htbp]
\includegraphics[width=\linewidth,angle=0]{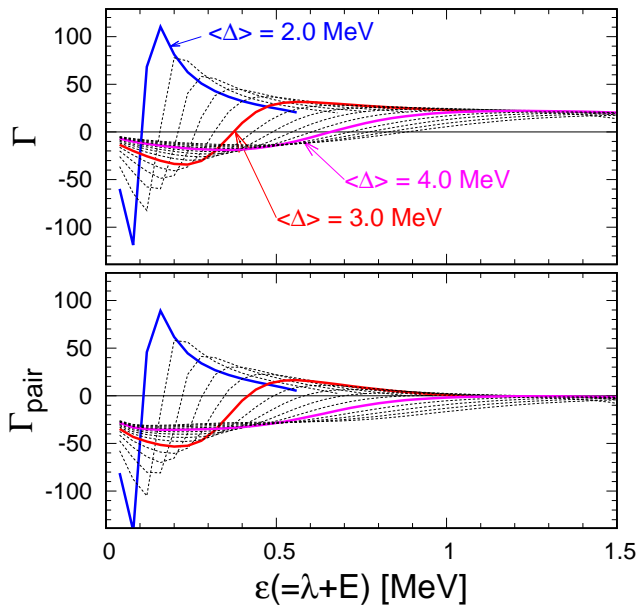}
\caption{(Color online) The pairing dependence of the circulations
  $\Gamma$ and $\Gamma_{pair}$ around the energy region for the p-type
  $p_{3/2}$ resonance. The solid blue, red and purple curves represent
  the circulation with $\bra\Delta\ket=2.0$, $3.0$ and $4.0$ MeV,
  respectively. The dotted curves shows the circulation calculated every
  $0.2$ MeV.}
\label{fig5}
\end{figure}

Fig.\ref{fig5} shows how the circulations $\Gamma$ and $\Gamma_{pair}$ change when the pairing gap
is varied around the energy region of the p-type $p_{3/2}$ quasiparticle resonance.
From the analysis of this figure, we found that the energy for which $\Gamma=0$ is found
almost at the same energy of the real part of the S-matrix pole, and the energy for which
$\Gamma_{pair}=0$ is found almost at the same energy with the K-matrix pole when
$\bra\Delta\ket<\bra\Delta\ket_c$, but no longer exists when $\bra\Delta\ket>\bra\Delta\ket_c$.

\section{Summary}

In this paper, we analyze the effect of pairing on incident neutron current in $n$-$A$
scattering described within the framework of HFB theory. For this purpose, the current was defined
using the HFB scattering wave function and its numerical calculation was performed.
To analyze the effect of pairing, the current was decomposed into HF part and pairing part,
and the corresponding vorticity and circulation were also calculated and analyzed.

It was found that the pairing effects on current, vorticity, and circulation are
quite different between p-type and h-type quasiparticle resonances.
In the case of the h-type quasiparticle resonances, the pairing has the effect of preventing the neutron current
from flowing into the nucleus, which reduces vorticity and circulation.
In the case of the p-type quasiparticle resonance, the pairing acts to prevent neutron current
from entering the nucleus in the energy region below the resonance energy, while the pairing acts
to allow neutron current from the backward region to enter in the energy region above the resonance
energy.
Vortices appear due to the presence of metastable structures of a certain partial wave component,
but the direction of the vortices is reversed before and after the resonance energy.
These features are consistent with the results of Eq.(\ref{approx-omega}), which shows how the
S-matrix poles of p-type and h-type quasiparticle resonances contribute to the vorticity.

We also found that the effect of the disappearance of the K-matrix pole due to the correlation
between the independent K-matrix pole near the quasiparticle resonance and the K-matrix of the
quasiparticle resonance (which has been discussed in Ref.\cite{classres}) appears in the form of
the disappearance of the energy which satisfies $\Gamma_{pair}=0$.

In order to discuss the dynamic features and properties of vortices in quantum systems in more
detail, it is necessary to derive the fluid equation based on the Schr\"{o}dinger
equation (including
the HFB equation, etc.) and discuss it in terms of the quantum pressure term that will appear
in the fluid equation, touching on the violation of Kelvin's circulation theorem and the
hydrodynamic conservation law.
However, we leave it for future work without touching on it in this paper.

\section*{Acknowledgments}
This work is funded by Vietnam National Foundation for Science and Technology
Development (NAFOSTED) under grant number “103.04-2019.329”.
T. V. Nhan Hao acknowledges the partial support of the Hue University under the Core
Research Program, Grant No. NCM.DHH.2018.09. Hoang Dai Nghia is funded by Vingroup JSC
and supported by the Master, PhD Scholarship Programme of Vingroup Innovation Foundation
(VINIF), Institute of Big Data, code VINIF.2021.ThS.32.

\end{document}